\documentclass[fleqn,usenatbib]{mnras}

\usepackage{newtxtext,newtxmath,cuted,lipsum}
\usepackage[T1]{fontenc} 
\usepackage{graphicx}	
\usepackage{subcaption}  
\usepackage{amsmath}	
\usepackage{amsfonts}
\usepackage{enumerate} 
\usepackage{colordvi} 
\usepackage{bm}
\usepackage{epsfig}
\usepackage{float}
\usepackage{verbatim}
\usepackage[dvipsnames]{xcolor}
\usepackage{multirow,multicol}
\usepackage{ulem}
\usepackage{hyperref}
\usepackage{fontawesome5} 
\usepackage{cleveref}

\DeclareMathAlphabet{\pazocal}{OMS}{zplm}{m}{n}
\newcommand{\unif}{\pazocal{U}}

\newcommand{\github}[1]{%
   \href{#1}{\faGithub}%
}

\newcommand{\code}[1]{{\texttt{#1}}}

\newcommand{\be}{\begin{equation}}
\newcommand{\ee}{\end{equation}}
\newcommand{\bea}{\begin{eqnarray}}
\newcommand{\eea}{\end{eqnarray}}

\title[KiDS-1000: Dark Scattering]{Dark Scattering: accelerated constraints from KiDS-1000 with \texttt{ReACT} and \texttt{CosmoPower}}  

\author[K. Carrion]{Karim Carrion,$^{1}$\thanks{E-mail: kcarrion@icf.unam.mx}
Pedro Carrilho,$^{2}$
Alessio Spurio Mancini,$^{3,4,5}$
Alkistis Pourtsidou,$^{2,6}$
Juan Carlos Hidalgo$^1$
\\
$^{1}$ Instituto de Ciencias F\'{i}sicas, Universidad Nacional Aut\'{o}noma de M\'{e}xico,
Cuernavaca, Morelos, 62210, Mexico\\ 
$^{2}$ Institute for Astronomy, The University of Edinburgh, Royal Observatory, Edinburgh EH9 3HJ, UK \\ 
$^{3}$ Department of Physics, Royal Holloway, University of London, Egham Hill, Egham TW20 0EX, UK \\ 
$^{4}$ Mullard Space Science Laboratory, University College London, Holmbury St. Mary, Dorking RH5 6NT, UK \\
$^{5}$ Department of Physics and Astronomy, University College London,
Gower Street, London WC1E 6BT, UK \\
$^{6}$ Higgs Centre for Theoretical Physics, School of Physics and Astronomy, The University of Edinburgh, Edinburgh EH9 3FD, UK
}
\date{Accepted XXX. Received YYY; in original form ZZZ}

\pubyear{2024}

\begin{document}
\label{firstpage}
\pagerange{\pageref{firstpage}--\pageref{lastpage}}
\maketitle

\begin{abstract}
We present constraints on the Dark Scattering model through cosmic shear measurements from the Kilo Degree Survey (KiDS-1000), using an accelerated pipeline with novel emulators produced with \code{CosmoPower}. Our main emulator, for the Dark Scattering non-linear matter power spectrum, is trained on predictions from the halo model reaction framework, previously validated against simulations. Additionally, we include the effects of baryonic feedback from \code{HMcode2016}, whose contribution is also emulated.
We analyse the complete set of statistics of KiDS-1000, namely Band Powers, COSEBIs and Correlation Functions, for Dark Scattering in two distinct cases. In the first case, taking into account only KiDS cosmic shear data, we constrain the amplitude of the dark energy -- dark matter interaction to be $\vert A_{\rm ds} \vert \lesssim 20$ $\rm b/GeV$ at 68\% C.L. 
Furthermore, we add information from the cosmic microwave background (CMB) from Planck, along with baryon acoustic oscillations (BAO) from 6dFGS, SDSS and BOSS, approximating a combined weak lensing + CMB + BAO analysis. From this combination, we constrain $A_{\rm ds} = 10.6^{+4.5}_{-7.3}$ $\rm b/GeV$ at 68\% C.L. We confirm that with this estimated value of $A_{\rm ds}$ the interacting model considered in this work offers a promising alternative to solve the $S_8$ tension. 
\end{abstract}


\section{Introduction}

Over the coming years cosmology will be transformed, with enormous amounts of new data being collected by Stage IV surveys such as \href{https://www.desi.lbl.gov}{\textbf{DESI}}\footnote{\url{https://www.desi.lbl.gov}}, which recently released its first set of data \citep{DESI:2023ytc}, the recently launched \href{https://www.euclid-ec.org}{\textbf{Euclid}}\footnote{\url{https://www.euclid-ec.org}} satellite mission \citep{laureijs2011euclid}, \href{https://www.lsst.org/about}{\textbf{The Vera Rubin Observatory}}\footnote{\url{https://www.lsst.org/about}} \citep{LSSTDarkEnergyScience:2012kar}, and the  \href{https://www.skatelescope.org/science}{\textbf{SKA Observatory}}\footnote{\url{https://www.skatelescope.org/science}} \citep{SKA:2018ckk}. These are going to observe several millions of galaxies over a large fraction of the sky, in the optical and radio wavelengths out to high redshifts. With these instruments it will soon be possible to probe scales approaching the size of the Hubble horizon, and determine with greater accuracy the properties of the Universe.

The standard cosmological model is $\Lambda$CDM, consisting of dark energy in the form of a cosmological constant $\Lambda$ and cold dark matter. Stage IV optical galaxy surveys will use galaxy clustering and weak gravitational lensing to probe the nature of dark energy and aim to provide sub-percent constraints on the equation of state of dark energy ($w$).

In recent years, as more data has become available, the $\Lambda$CDM model has come into question due to apparent inconsistencies (tensions) between observables. Firstly, the $H_0$ measurements from early and late-time probes differ significantly, at a level of $\sim 5\sigma$ \citep{Riess:2019qba, DiValentino:2020zio}. Secondly, the amplitude of density fluctuations at early times, encoded in the $\sigma_8$ value extrapolated to the present day, is inconsistent with the near redshift measurements of that same quantity (see \cite{Douspis:2018xlj}). 
This second tension is most evident in the inferred values of the parameter $S_8 \equiv \sigma_8 \sqrt{\Omega_{\rm m}/0.3}$: from the latest cosmic microwave background (CMB) measurements of the Planck satellite we have $S_8 = 0.825 \pm 0.011$ \citep{Planck:2018vyg}, whereas at low redshifts, weak lensing measurements from the DES collaboration \citep{PhysRevD.105.023520} have estimated $S_8 = 0.776 \pm 0.017$ and the Kilo-Degree Survey (KiDS-1000) \citep{KiDS:2020suj} reports a value of $S_8 = 0.759^{+0.024}_{-0.021}$. Furthermore, the joint analysis of these last two surveys \citep{Kilo-DegreeSurvey:2023gfr} found a value of $S_8 = 0.790^{+0.018}_{-0.014}$. Thus, discrepancies are around $2 - 2.5 \sigma$ with respect to Planck. While such discrepancies may be attributed to systematics or unaccounted astrophysical processes, the role of cosmological modelling cannot be entirely ruled out \citep{Cortes:2023dij}.

In the pursuit of alternatives to the concordance model, interacting dark energy (IDE) models stand out as potential candidates. Such models can include a variety of interactions between dark matter and dark energy. A few IDE models have been shown to be able to mitigate the existing tensions through various observational probes, including CMB \citep{Zhai:2023yny}, galaxy clustering \citep{Carrilho:2023qhq}, weak lensing \citep{An:2017crg}, supernovae observations \citep{DiValentino:2019ffd,Gao:2021xnk,Yao:2022kub}, and gravitational waves (GW) \citep{Li:2019ajo}. Of particular interest is the interaction characterised by pure momentum transfer between dark energy and dark matter, which is able to alleviate the $S_8$ tension. A specific model with pure momentum transfer is called Dark Scattering (DS). The model takes inspiration from Thomson scattering and postulates an elastic interaction between the two dark sector species. In this paper we will focus on constraining this model using KiDS data, following our previous work on validation with simulations \citep{Carrilho:2021rqo}, which we denote hereafter as ``P1''.
 
\begin{figure*} 
  \begin{subfigure}{\columnwidth}
    \centering
    \includegraphics[width=\textwidth]{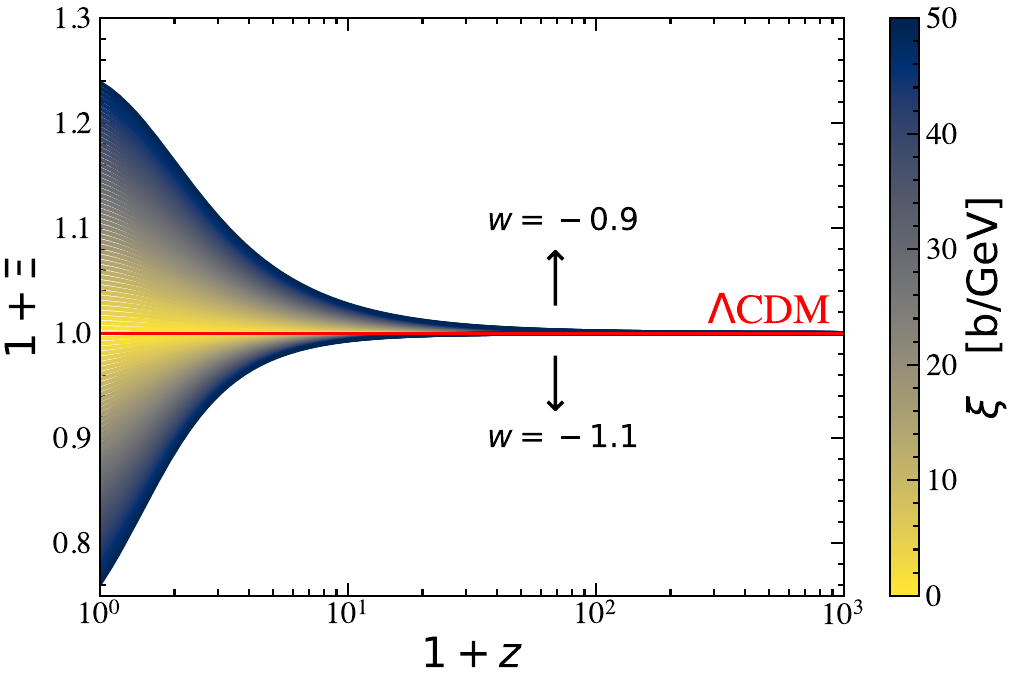}
    \caption[]%
            {}
    \label{Fig:interaction_term_a}
  \end{subfigure}
  \begin{subfigure}{\columnwidth}
    \centering
    \includegraphics[width=\textwidth]{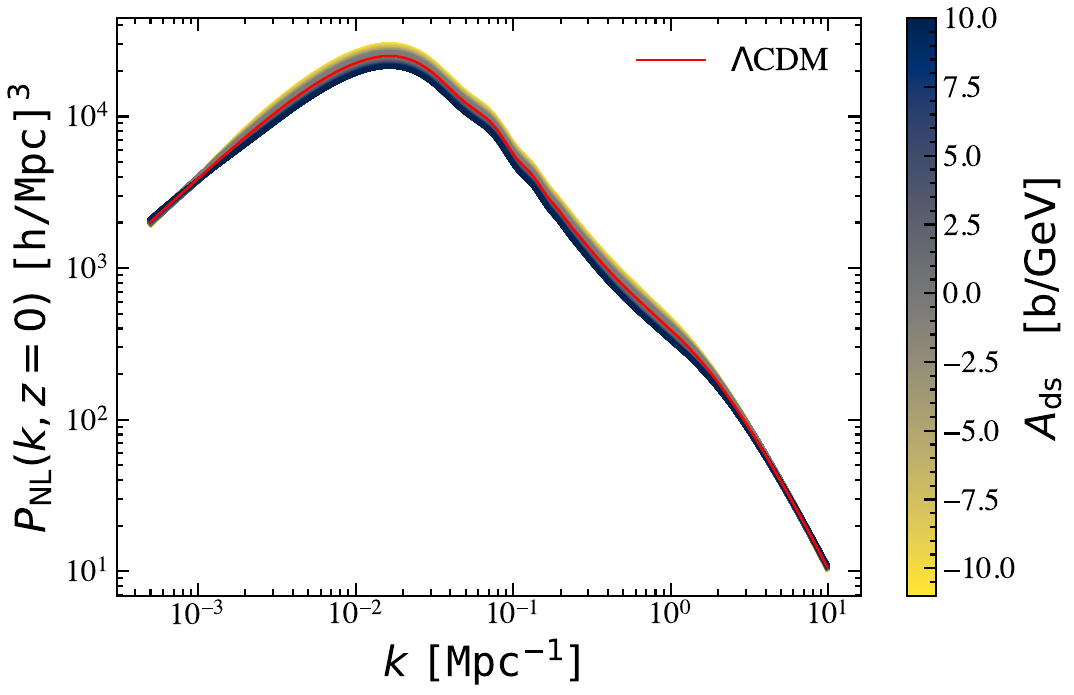}
    \caption[]%
            {}
    \label{Fig:interaction_term_b}
  \end{subfigure}
    \caption[Behavior of the DS interaction term]{a) Interaction term $\Xi$ as a function of redshift, for a range of constant equation of state parameter values. b) Depending on the sign of the interaction parameter $A_{\rm ds}$, we get an enhancement or suppression of the matter power spectrum amplitude.}
\label{Fig:interaction_term}
\end{figure*}

To speed-up the inference pipeline, we have trained and incorporated new Neural Network (NN) emulators, elaborating on previous work by \cite{SpurioMancini2022} (see also e.g. \cite{Arico:2021izc,Mootoovaloo:2021rot,Gunther:2022pto,Bonici:2022xlo,Nygaard:2022wri,Piras:2023aub} for similar efforts). From a computational standpoint, emulators have consistently demonstrated significantly higher speed in model processing when compared to Boltzmann codes like \code{CLASS} \citep{Lesgourgues2011} or \code{CAMB} \citep{Lewis:1999bs}, while leading to essentially the same parameter constraints at a fraction of the time.

In particular we will take advantage of the publicly available code called \code{CosmoPower} \citep{SpurioMancini2022} to create emulators of power spectra, which notably accelerate Bayesian inference process by orders-of-magnitude, as exemplified in recent analyses of CMB data \citep{SPT-3G:2022hvq, Bolliet:2023sst, Balkenhol:2024sbv}, and weak lensing analyses for interacting dark energy and modified gravity models \citep{ManciniSpurio:2021jvx, SpurioMancini:2023mpt}. 

\noindent The paper is organised as follows: In \autoref{sec:models}, we briefly review IDE theories and introduce the DS model. In \autoref{subsec:reaction} we briefly review the halo model reaction formalism. Following the results of P1, we compute the DS matter power spectrum using the \code{ReACT} code. In \autoref{sec:emulation} we present the pipeline for developing our emulators based on \code{CosmoPower}. We report the accuracy of each emulator as well as its computational efficiency. In \autoref{sec:data_setup} we briefly describe the KiDS-1000 weak lensing data and the setup of the Bayesian analysis. Our results are presented in \autoref{sec:results}. We first present the KiDS-1000 analysis and report the derived constraints in \autoref{subsec:free_cmb}. In \autoref{subsec:with_cmb}, we also include a combined KiDS+CMB+BAO inference. 
Lastly, we summarize and draw conclusions in \autoref{sec:conclusion}. 

\section{Model}\label{sec:models}
\subsection{The Dark Scattering model}

The nature of dark matter and dark energy is currently unknown and the fact that they are considered uncoupled in $\Lambda$CDM is an assumption.
This has motivated the development of interacting dark energy (IDE) models including a (non-gravitational) coupling current $Q^\nu$ that represents the energy and momentum exchange between dark matter and dark energy (see \citet{Pourtsidou:2013nha} for a general approach using the pull-back formalism for fluids). In the presence of this interaction, the energy-momentum tensors of dark matter and dark energy are no longer separately conserved. That is
\bea
\nabla_\mu T^{\mu \nu}_{\rm c} = Q^\nu, \quad \Longleftrightarrow \quad \nabla_\mu T^{\mu \nu}_{\rm DE} = -Q^\nu.
\label{Eq:IDE_coupled}
\eea

\begin{figure}
\centering
 \begin{subfigure}[t]{0.455\textwidth}
\includegraphics[width=\textwidth]{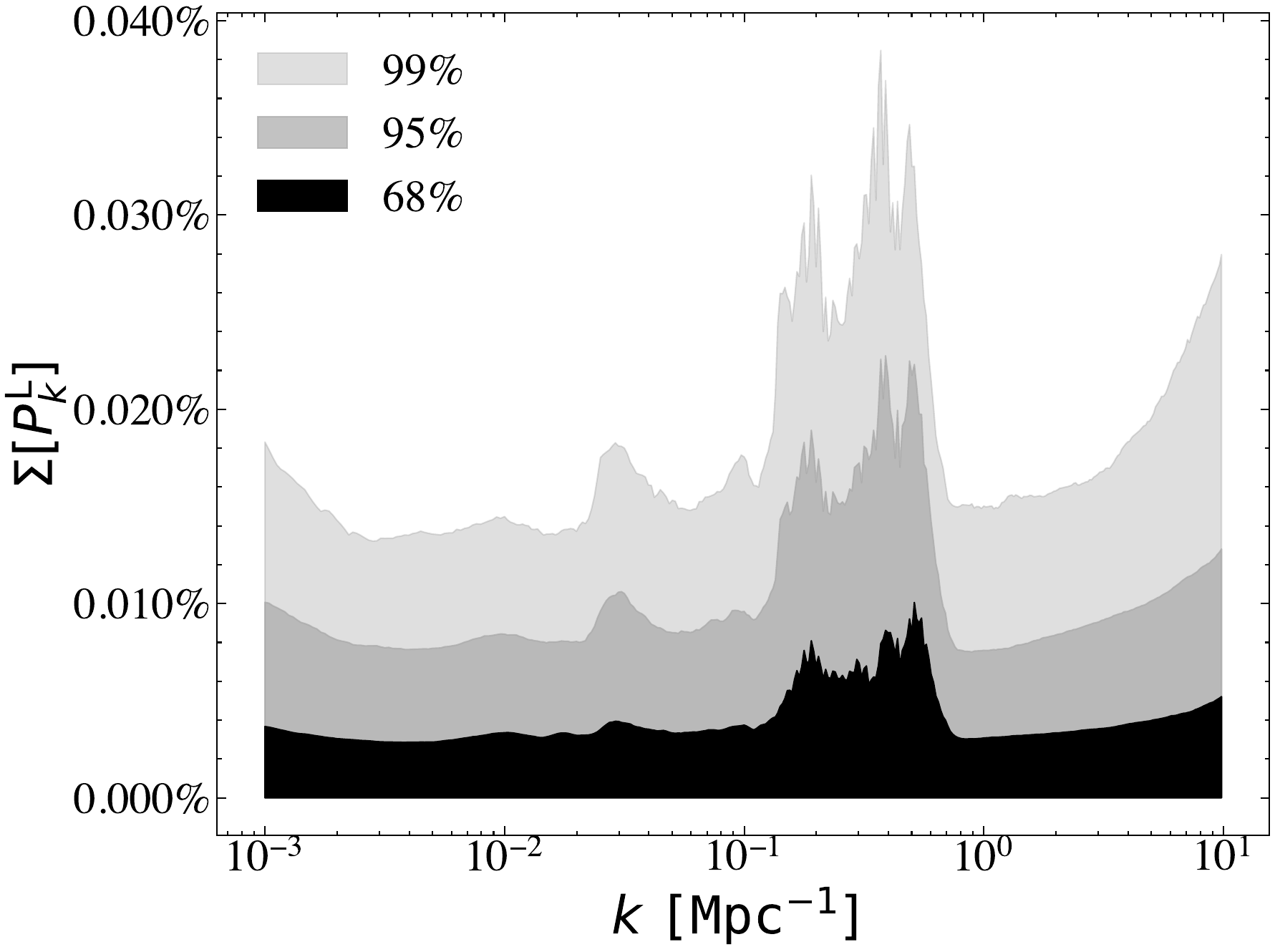}
    \caption[]%
            {}{}
\label{fig:timing1}
\end{subfigure}
\centering
\begin{subfigure}[t]{0.455\textwidth}
\includegraphics[width=\textwidth]{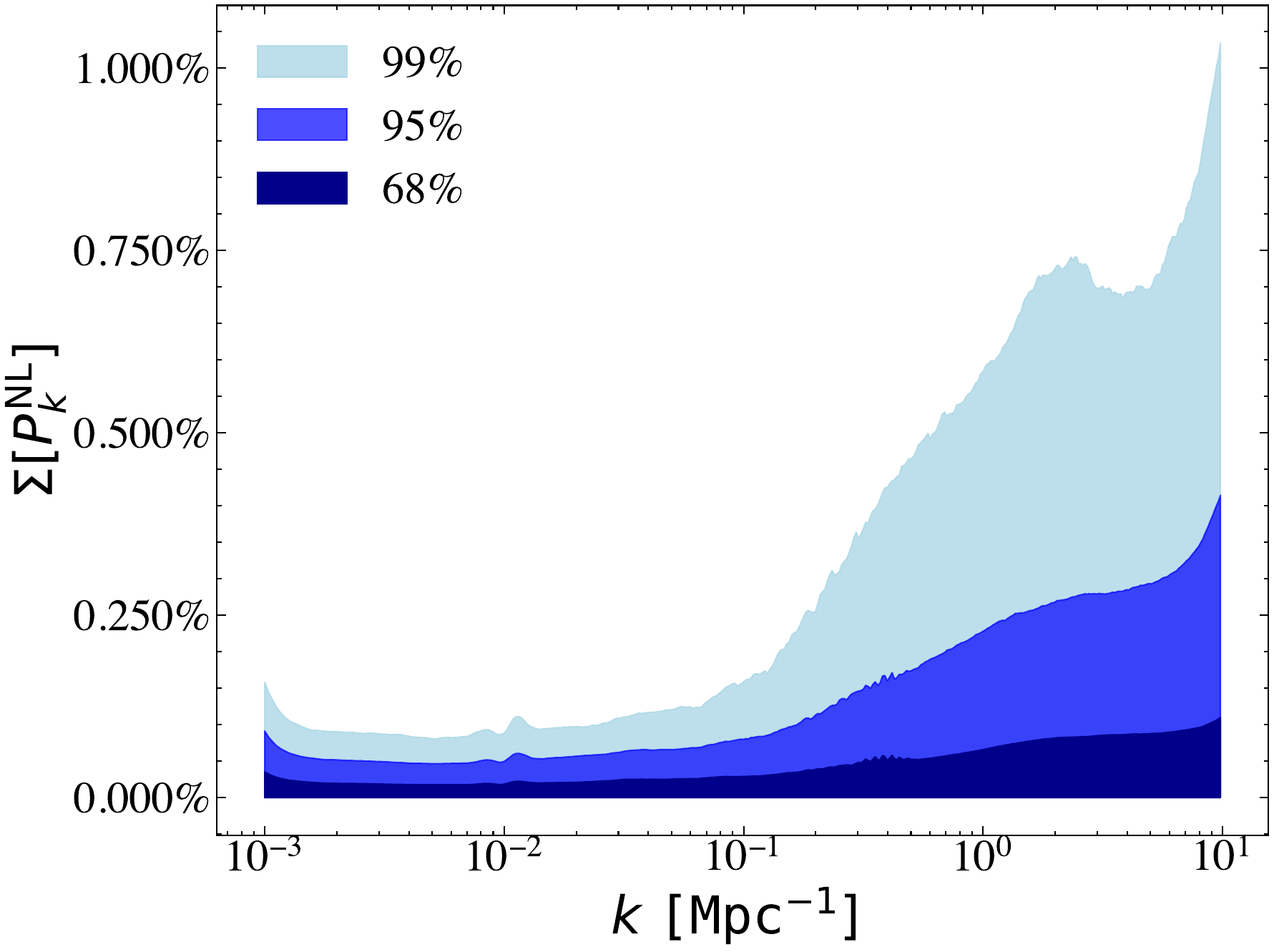}
    \caption[]%
            {}{}
\label{fig:timing2}
\end{subfigure}
\centering
\begin{subfigure}[t]{0.455\textwidth}
\includegraphics[width=\textwidth]{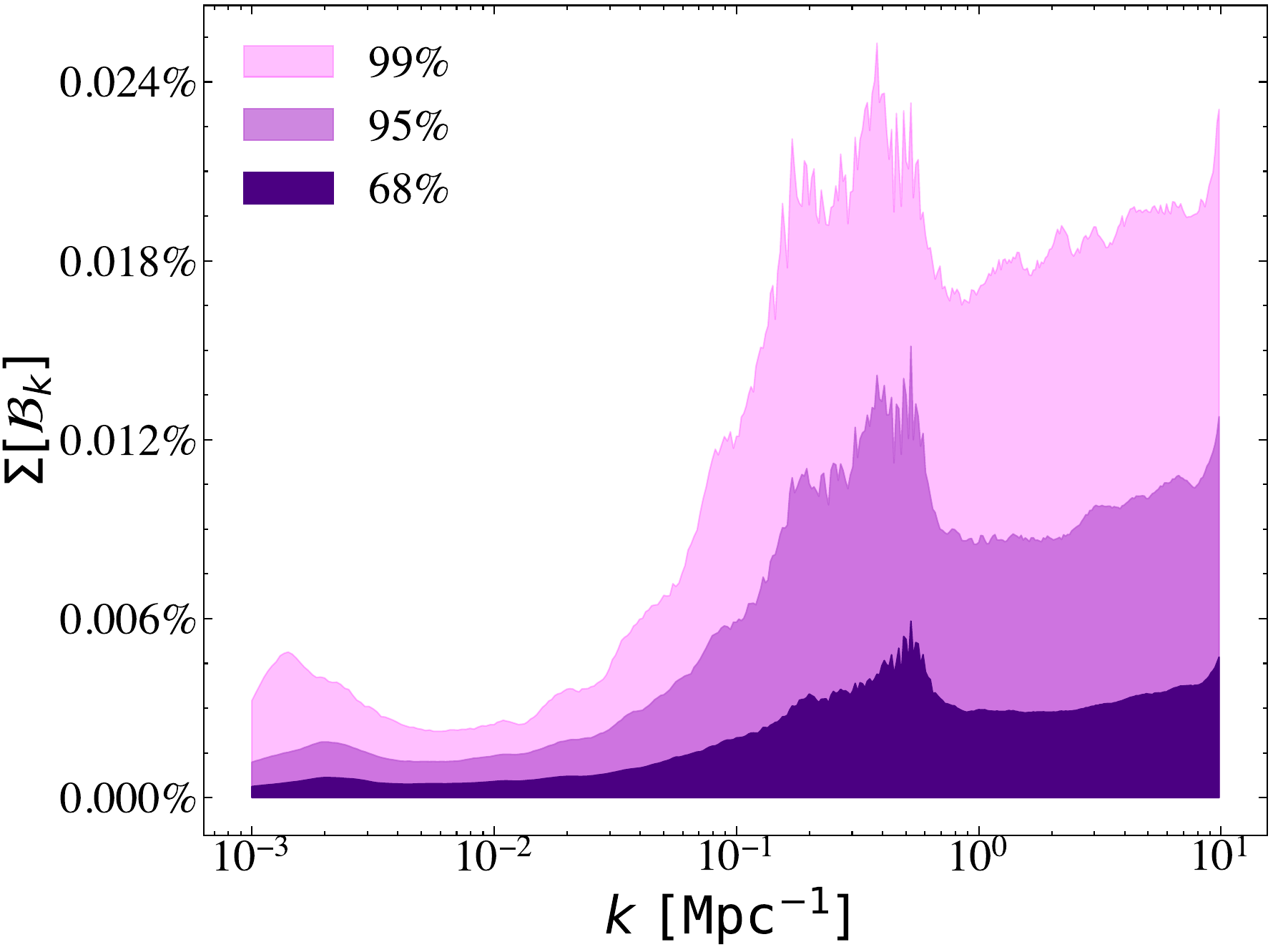} 
    \caption[]%
            {}{}
\label{fig:timing3}
 \end{subfigure}
\caption{Comparison of our emulators against $\sim 10^4$ spectra from the testing set, dubbed ``real''. The percentage absolute emulator error is calculated as $\Sigma[F_k] = 100\,\mathcal{\%} \cdot \left| \frac{ F_{k, \rm{emulated}} - F_{k, \rm{real}}} {F_{k, \rm{real}}}\right|$, with $F_{k} = \{P_{k}^{\rm{L}}, P_{k}^{\rm{NL}}, \mathcal{B}_{k}\}$, and each case has enclosed the areas of the 68, 95 and 99 percentiles. (a) The linear emulator in the black, grey and light grey enclose the percentiles of $\Sigma[P_{k}^{\rm{L}}]$. (b) Likewise, the Dark Scattering DM-only non-linear spectrum in the dark blue, blue and cyan are for the percentiles of $\Sigma[P_{k}^{\rm{NL}}]$. (c) Whereas the baryonic boost of \autoref{Eq:boost_eq} in the dark purple, violet and pink for the percentiles of $\Sigma[\mathcal{B}_{k}]$. Note a reduction in accuracy plots appears at BAO wiggles when $k\sim [0.1, 1] \ \rm{Mpc}^{-1}$ due to the sensitivity of cosmological parameters on this region.}
\label{Fig:accuracy_emu}
\end{figure}

In this paper we will investigate a pure momentum exchange model known as Dark Scattering (DS), initially proposed by \citet{Simpson:2010}. The coupling is given by  
\bea
 \mathbf{Q}_{{\rm DS}}=  -(1+w) \sigma_{\rm DS} \rho_{\rm DE} n_{\rm c}  (\mathbf{v}_{{\rm c}}-\mathbf{v}_{{\rm DE}}) \, ,
 \label{Eq:DS_eq}
\eea
where $\mathbf{v}$ is the fluid velocity, the interaction cross section is denoted by $\sigma_{\rm DS}$ and number density of CDM particles is $n_{c}$. While the model is valid for a general evolution of the equation of state parameter, $w$, we will take it to be constant in this work.

Subsequently, in \citet{Baldi:2014} and \citet{Baldi:2016zom} the DS effects were studied through full N-body simulations, thereby exploring the non-linear physics of this cosmology. In those simulations, the effects of DS were applied to the equation of motion of particles as, 
\bea
\dot{\mathbf{u}}_c = - (1+\Xi) H \mathbf{u_c} - \nabla_\mathbf{r} \Phi.
\label{Eq:interaction_eom}
\eea
Here, $H=\dot a/a$ is the Hubble rate, $\mathbf{u}$ is comoving particle velocity, $\Phi$ is the gravitational potential and we define the interaction term as follows,
\bea
\Xi(z) \equiv   \xi \left(1+w\right) \dfrac{3\Omega_{\rm DE}}{8\pi G} H \, ,
\label{Eq:Interaction_term}
\eea
As illustrated in P1, the equation of motion \autoref{Eq:interaction_eom} is also reached at the Newtonian limit of the full model. The above term depends only on background quantities and the coupling strength parameter $\xi \equiv \sigma_{\rm DS}/m_{c} \geq 0$ in units $\rm [b/GeV]$. In addition, the DS model can be considered as an extension of $w$CDM. It has a well-defined $\Lambda$CDM limit when $w \rightarrow -1$, which gives $\Xi=0$ (see \autoref{Fig:interaction_term_a}). 

In the case where massive neutrinos are included, the effective strength of the interaction is modulated by $f_c=\rho_c/\rho_{\rm m}$ the dark matter fraction of the total matter (see P1 for details), so that we can define an effective coupling constant $\bar\xi$, given by
\bea
\bar\xi=\frac{f_c}{1+\Xi_0(1-f_c)}\xi\,,
\label{eff_xi}
\eea
with $\Xi_0\equiv\Xi(z=0)$.
 
Thus, we define the effective interaction amplitude as, 
\bea
A_{\rm ds} \equiv   \bar\xi \left(1+w\right). 
\label{Eq:Interaction_term_eff}
\eea
In \autoref{Eq:interaction_eom} the effect of the interaction manifests as an additional frictional force of dragging or pushing on the CDM particles. As expected, this alters the matter power spectrum depending on ${\rm{sign}}(A_{\rm ds})$, as illustrated in  \autoref{Fig:interaction_term_b}. 
The main aim of this paper is to test the DS model for the first time with a weak lensing probe (the KiDS-1000 dataset) and robust non-linear modelling.

\subsection{The halo model reaction}\label{subsec:reaction}

We model the non-linear matter power spectrum using an adapted halo model framework, an extension of the conventional halo model \citep{Cooray:2002dia,Asgari:2023mej} applied to beyond-$\Lambda$CDM (alternative) models \citep{Cataneo:2018cic,Bose:2020wch,Cataneo:2022xvp}. The theory modifications are encoded in a function called reaction, $\mathcal{R}(k,z)$, which is defined as  
\bea
 P^{\rm alt}_{\rm NL}(k,z) \equiv \mathcal{R}(k,z) \times   P_{\rm NL}^{\rm pseudo}(k,z) \, .
 \label{Eq:Reaction_spectrum}
\eea
The `alt' subscript stands for alternative theory, while the `pseudo' term refers to a $\Lambda$CDM cosmology with initial conditions matched so that $P^{\rm alt}_{\mathrm{L}}(k,z) = P^{\rm pseudo}_{\mathrm{L}}(k,z)$ at the target redshift $z$. The reaction function may account for modifications to the theory of gravity, massive neutrino contributions or a non-standard dark sector, included in the following form:
\bea
    \mathcal{R}(k,z)=\frac{\bar{f}_{\nu}^{2} P_{\mathrm{HM}}^{(\mathrm{cb})}(k,z)+2 f_{\nu}\bar{f}_{\nu} P_{\mathrm{HM}}^{(\mathrm{cb} \nu)}(k,z)+f_{\nu}^{2} P_{\mathrm{L}}^{(\nu)}(k,z)}{P_{\mathrm{L}}^{(\mathrm{m})}(k,z)+P_{\mathrm{1h}}^{\mathrm{pseudo}}(k,z)}  \, .
    \label{eq:reaction_def}
\eea
The superscript $\rm (m) \equiv (cb+\nu) $ accounts for the sum of matter components, with `cb' standing for CDM plus baryons and `$\nu$' for massive neutrinos. Here $\bar{f}_{\nu} = 1 - f_{\nu}$ with $f_\nu = \rho_\nu/ \rho_\mathrm{m}$. The subscripts `L' refer to linear and `1h' to 1-halo term. \citet{Bose:2020wch} have included the reaction formalism in a publicly available \texttt{C++} code, called \code{ReACT}\footnote{We refer the reader to \cite{Cataneo:2018cic}, \cite{Bose:2020wch}, \cite{Bose:2021mkz}, \cite{Carrilho:2021rqo}, \cite{Cataneo:2022xvp} and \cite{Bose:2022vwi} for halo model reaction applications and validation against simulations.}. The code has the advantage of including a Python wrapper, which allows us to run \code{ReACT} within a Python interpreter. In P1, the DS formulation for \code{ReACT} was implemented and validated for several values of the coupling parameter $\xi$ against N-body simulations. 

\section{EMULATORS}
\label{sec:emulation} 

\subsection{Emulating the matter power spectrum}

In light of the detailed validation of the DS halo model predictions against simulations in P1, in this work we proceed to produce two accurate and fast DS matter power spectrum emulators with the aid of \code{CosmoPower} \citep{SpurioMancini2022}. The NN architecture and values of hyperparameters are preserved from the original paper.
In order to generate the training set of non-linear and linear spectra, we produce a set of spectra for a range of values of 8 cosmological parameters plus a given redshift $z$, as follows: 
\bea 
\theta_{\rm DS} = \{\omega_{\rm b}, \omega_{\rm cdm}, h, n_s, S_8, m_\nu, w, A_{\rm ds}, z \},
\label{Eq:DS_parameters_training}
\eea
where $\omega_i = \Omega_i h^2$ is the physical density parameter, $h$ is the Hubble parameter, $n_s$ is the scalar spectral index and $m_\nu$ is the neutrino mass. For our statistical analysis, it is an advantage to employ $S_8$ as an input parameter of the emulators instead of $\ln(10^{10} A_s)$. This is motivated by the fact that weak lensing measurements are more sensitive to $S_8$, and also because our reaction-based modelling is directly dependent on the late-time amplitude, as explained below. We proceed according to the following pipeline: Initially, we take the parameter set in \autoref{Eq:DS_parameters_training} to calculate the linear power spectrum from the modified version of \code{CLASS} presented in P1. This is stored to train the linear emulator. Subsequently, we derive the reaction described in \autoref{eq:reaction_def} using \code{ReACT} and the linear spectrum as an input. Since the pseudo spectrum adheres to the standard halo model approach, we opt to use \code{HMcode2020} \citep{Mead:2020vgs}. This choice is motivated by its capability to cover a wide cosmological parameter range, in contrast to alternatives such as \code{EuclidEmulator2} \citep{Euclid:2020rfv}, which are bound by more restricted parameter ranges. Finally, we take the product of the pseudo spectrum times the reaction in order to compute the DS non-linear power spectrum.

It is important to emphasize that the validity range of the emulator must be carefully determined in agreement with the validity range of the codes, as well as in concordance with the data to be analysed. 
This step is crucial in order to prevent emulator re-training stages. In our case, the parameter range is limited to those values where \code{ReACT} can resolve the halo model spherical collapse. Specifically, we are free to choose values for a set of cosmological parameters which yield $\sigma_8(z=0)$ values between $0.55$ and $1.4$. In addition, the validity range must be consistent with KiDS-1000 official $\Lambda$CDM analysis.
The range of our set of parameters is shown in \autoref{tab:range_of_emulators}. Within that range we generated $\sim 5\cdot 10^5$ spectra for the training set and put aside $10 \%$ for the testing set. In the training stage, the $k$-modes in the emulators are restricted to the range $[10^{-3}, 10]\ \text{$1/$Mpc}$, while the redshift $z$ is traced up from 0 to 5 and treated as an additional input parameter for both emulators as shown in \autoref{tab:range_of_emulators}. 

\subsection{Emulating the baryonic feedback}

The contribution of baryonic effects on the matter power spectrum \citep{Chisari:2019tus,Giri:2021qin,Arico:2023ocu} is also taken into account through a baryonic factor $\mathcal{B}(k,z)$, defined as the ratio between the full power spectrum and the DM-only spectrum, as follows, 
\bea
\mathcal{B}(k,z) = \dfrac{P_{\rm full}}{P_{\rm DM\text{-}only}} \, .
\label{Eq:boost_eq}
\eea
\noindent This is computed from a fitted model in terms of two baryonic parameters $c_{\rm min}$ and $\eta_0$, which captures the influence of baryons within a halo. Here, $\eta_0$ is determined by the formula $\eta_0 = 0.98 - 0.12 c_{\rm min}$ but can alternatively be treated as independent in our emulator. The case where baryons have no influence corresponds to the formula above with a value $c_{\rm min} = 3.13$ (DM-only). 

To proceed we have produced $\sim 10^5$ values of the baryonic factor using \code{HMCODE2016}\footnote{Our choice of \code{HMCODE2020} over \code{HMCODE2016} for computing the DM-only non-linear power spectrum is driven by its improved modelling of BAO damping and an updated treatment of massive neutrinos \citep{Mead:2020vgs}. However, we choose instead \code{HMCODE2016} for baryonic feedback because it has a DM-only limit where baryonic effects vanish, unlike the 2020 version.} \citep{Mead:2015yca} which we store in a training set in order to emulate \autoref{Eq:boost_eq} (keeping once again 10$\%$ of those for testing). This baryonic feedback emulator contains the following parameters:
\bea
\theta_{\rm feedback} = \{\omega_{\rm b}, \omega_{\rm cdm}, h, n_s, S_8, c_{\rm min}, \eta_0 ,z \}.
\label{Eq:boost_parameters_training}
\eea
\noindent We consider the same ranges for the $k$-modes. The redshift is also varied from 0 to 5. Notice this baryonic feedback emulator incorporates fewer input parameters (specifically, only $\Lambda$CDM ones) than the DS emulator due to the fact that baryonic feedback is expected to be accurate enough for alternatives to $\Lambda$CDM, and is weakly affected by most cosmological parameters (with the exception of the baryon fraction \citep{Angulo:2020vky, Giri:2021qin,Arico:2020lhq}).

\begin{table}
  \centering 
  \begin{tabular}{|c|c|}
    \textbf{Parameter} & \textbf{Validity}  \\ \hline \hline $\omega_{\mathrm{b}}$   &  [0.01865, 0.02625] \\ \hline 
    $\omega_{\mathrm{cdm}}$ &  [0.05,    0.255] \\ \hline
    $h$  & [0.64,    0.82] \\ \hline
    $n_s$ & [0.84,    1.1]  \\ \hline
    $S_8$ & [0.6, 0.9] \\ \hline
    $w$  & [-1.3, -0.7]   \\ \hline
    $m_\nu$  & [0,    0.2] eV   \\ \hline
    $|A_{\rm ds}|$   &  [0, 30]  \rm b/GeV \\ \hline
    $z$   &  [0, 5] \\ \hline
    $c_{\rm min}$ & [2.0, 4.0]  \\ \hline
    $\eta_0$ & [0.5, 1.0]\\ 
    \hline
    \hline
    \end{tabular}
   \caption{Input parameters and their validity range for the emulators produced in this work. The spectra are produced in the range
   $-3 < \log_{10} (k\, \text{Mpc}) < 1$.}
\label{tab:range_of_emulators}
\end{table}

\subsection{Accuracy and efficiency of the Emulators}\label{subsec:accuracy}

We report the accuracy of our trained emulators in \autoref{Fig:accuracy_emu}. The top panel shows the accuracy of the DS linear power spectrum emulator with 99$\%$ of the testing set producing differences smaller than $0.05\%$ to the real value across the entire $k$-range considered, with a slight decrease in accuracy in the region corresponding to the BAO wiggles, which is more difficult to compute for the Boltzmann code, due to its sensitivity to cosmological parameters. Consequently, this difficulty is inherited to emulators during training stage.
The middle panel shows that the accuracy of the DS non-linear power spectrum emulator is better than $1\%$ up to $k = 10 \ h/\rm{Mpc}$, thus we are reproducing the output from \code{ReACT} with high precision. As already seen in \cite{SpurioMancini2022}, the differences between the emulated and the real predictions increase at highly non-linear scales. This reflects the intrinsic scatter in the real predictions arising from the numerical complexity of the computation performed by the modelling in that region. At the bottom panel we display the accuracy of the third emulator, with baryonic effects. In short, 99$\%$ are emulated with an error smaller than less than 0.03$\%$.

In terms of computational speed, there is already a major improvement at this stage. While employing \code{ReACT} to generate $\sim 2 \cdot 10^5$ samples required approximately 2 days of parallel processing across 100 CPU cores, the DS emulator accomplishes the same task in around $\sim 30$ seconds, in a single CPU core. 
For this reason, we replace the Boltzmann code calls for computing the matter power spectra within the KiDS-1000 pipeline with the DS emulator.

\section{DATASET AND SETUP}
\label{sec:data_setup} 

\subsection{KiDS-1000}

The data products we analyse are from the fourth data release of the Kilo-Degree Survey \citep{Kuijken:2019gsa,Wright:2019fwm,Hildebrandt:2020rno,Giblin:2020quj}, named KiDS-1000, which contains contains measurements of cosmic shear spanning spanning $\sim 1000 \ \text{deg}^2$. In this paper, we maintain the original cosmic shear and photometric redshift measurements, as well as the data modelling presented in preceding KiDS-1000 analyses \citep{Heymans:2020gsg,KiDS:2020ghu,Joachimi:2020abi}. Similarly to \citet{KiDS:2020suj}, we also consider three sets of statistics: Band Powers (BP, \cite{vanUitert:2017ieu}), real space two-point correlation functions (2PCFs, \cite{Schneider:2002jd}) and Complete Orthogonal Sets of E$/$B-Integrals (COSEBIs, \cite{Schneider2010}). We adopt the Non-Linear Alignment (NLA) model \citep{Hirata:2004gc,Bridle:2007ft} to describe the systematic effect of intrinsic alignments (IA) of galaxies with the surrounding matter distributions.

In order to compute the full non-linear power spectrum, we can achieve this by multiplying the emulated components as follows:
\bea
P^{\rm NL}_{\rm full}(k,z) =   \mathcal{B}(k,z) \times P^{\rm NL}_{\rm DM\text{-}only} (k,z) \, .
\label{full:spectrum}
\eea
Throughout our analysis, we assume two massless neutrinos and one massive neutrino with a mass fixed to $0.06$ eV. For the DS parameter $A_{\rm ds}$ and $w$, we assume flat prior distributions $|A_{\rm ds}| \rightarrow \unif [0.0, 30.0]$ b/GeV and $w \rightarrow \unif [-1.3, -0.7]$. Regarding the justification of the nuisance parameters (including the IA parameter) priors we refer to \citet{KiDS:2020suj}. 

We perform the Bayesian analysis by using \code{Montepython} \citep{Audren:2012wb} in which our emulators were internally implemented in the pipeline. In addition, we selected the sampler \code{Multinest} \citep{Feroz:2008xx} in order to also obtain the Bayes-factor values for each cosmology scenario. Lastly for this section, we share the main \code{Multinest} settings of our runs: \texttt{n\_live\_points = 1000}, \texttt{sampling\_efficiency = 0.3}, \texttt{n\_iter\_before\_update = 200}, \texttt{evidence\_tolerance = 0.01}, \texttt{boost\_posteriors = 10.0}, while the remaining parameters are set with default values from \code{MultiNest} itself.

\begin{table*}
\resizebox{\textwidth}{!}{%
\begin{tabular}{|c|ccc|ccc|ccc|}
  & \multicolumn{3}{c}{\textbf{Band Powers}}                       & \multicolumn{3}{c}{\textbf{COSEBIs}}                           & \multicolumn{3}{c}{\textbf{2PCFs}} \\
  \hline 
  & $\Lambda$CDM              & wCDM                       & DS
  & $\Lambda$CDM              & wCDM                       & DS
  & $\Lambda$CDM              & wCDM                       & DS \\
  \hline 
    $\Omega_{\mathrm{m}}$ & $0.328^{+0.073}_{-0.31}$ & $0.335^{+0.082}_{-0.11}$ &
    $0.353^{+0.092}_{-0.11}$ &  
    $0.292^{+0.06}_{-0.11}$ &  
    $0.293^{+0.064}_{-0.11}$ & 
    $0.293^{+0.066}_{-0.11}$ & 
    $0.228^{+0.035}_ {-0.06}$ & 
    $0.228^{+0.039}_{-0.063}$ &
    $0.226^{+0.038}_{-0.064}$ \\
    \hline
    $\sigma_8$ & $0.74^{+0.11}_{-0.15}$ & $0.74^{+0.1}_{-0.15}$ & 
    $0.708^{+0.087}_{-0.16}$ & 
    $0.790^{+0.13}_{-0.15}$  & 
    $0.792^{+0.12}_{-0.15}$ & 
    $0.79^{+0.13}_{-0.16}$ & 
    $0.90 \pm 0.1$ & 
    $0.903^{+0.098}_{-0.12}$ &
    $0.9^{+0.11}_{-0.12}$ \\
    \hline
    $S_{8}$ & 
    $0.752^{+0.031}_{-0.023}$ & 
    $0.754^{+0.034}_{-0.031}$ & 
    $0.739 \pm 0.036$ & 
    $0.751^{+0.026}_{-0.019}$ & 
    $0.753 \pm 0.029$ & 
    $0.750 \pm 0.031$ & 
    $0.766 \pm 0.019$ & 
    $0.770^{+0.025}_{-0.028}$ &
    $0.767^{+0.026}_{-0.031}$ \\
    \hline
    $w$ & 
    --- & 
    $-0.96^{+0.24}_{-0.13}$ & 
    $-0.99^{+0.2}_{-0.15}$ & 
    --- & 
    $-0.98^{+0.22}_{-0.14}$ & 
    $-1.05^{+0.1}_{-0.2}$ & 
    --- & 
    $-0.99^{+0.22}_{-0.14}$ &
    $-1.07^{+0.082}_{-0.19}$ \\ 
    \hline
    $A_{\rm ds}$ & 
    --- & 
    --- & 
    $-0.3^{+13}_{-8.5}$ & 
    --- & 
    --- & 
    $-3.6^{+7.8}_{-9.9}$ & 
    --- & 
    --- &
    $-4.8^{+7.1}_{-10}$ \\
    \hline
    $\log_{10} K$   &
    --- &
    $-0.1295 \pm 0.0011$ &
    $0.0787 \pm 0.0018$ &
    --- &
    $-0.1512 \pm 0.0003$  &
    $-0.3780 \pm 0.0021$    &
    --- &
    $-0.0761 \pm 0.0002 $  &
    $-0.4504 \pm 0.0027 $ 
    \end{tabular}
}
    \caption{Mean and marginalised 68$\%$ constraints on key weak lensing parameters from the KiDS-1000 analysis. We report the log-Bayes factors of each model with respect to $\Lambda$CDM with $K = \frac{\mathcal{Z}_{\rm{model}}}{\mathcal{Z}_{\Lambda\mathrm{CDM}}}$. According to Jeffreys' scale, a value of $|\log_{10} K|$ below 0.5 implies an indecisive advantage over $\Lambda$CDM. Note that the constraints on $w$ are prior dominated for all probes.}
    \label{tab:bestfit_K1K_params}
\end{table*}

\section{RESULTS}
\label{sec:results} 

\subsection{KiDS-1000 analysis}\label{subsec:free_cmb} 

The setup of priors on the cosmological parameters are limited to the validity range of the emulators, previously shown in \autoref{tab:range_of_emulators}. We assume flat distributions on these priors. Before producing posteriors for alternative models, and as a crosscheck, we used our non-linear spectrum + baryonic feedback emulators to reproduce the $\Lambda$CDM constraints obtained from the KiDS-1000 official results presented in \citet{KiDS:2020suj}. The comparison is presented in appendix \ref{app:A}. The contours are produced in around 10 minutes with \code{CosmoPower}, as opposed to the few days required with a Boltzmann solver.

\begin{figure} 
    \centering    \includegraphics[width=0.5\textwidth]{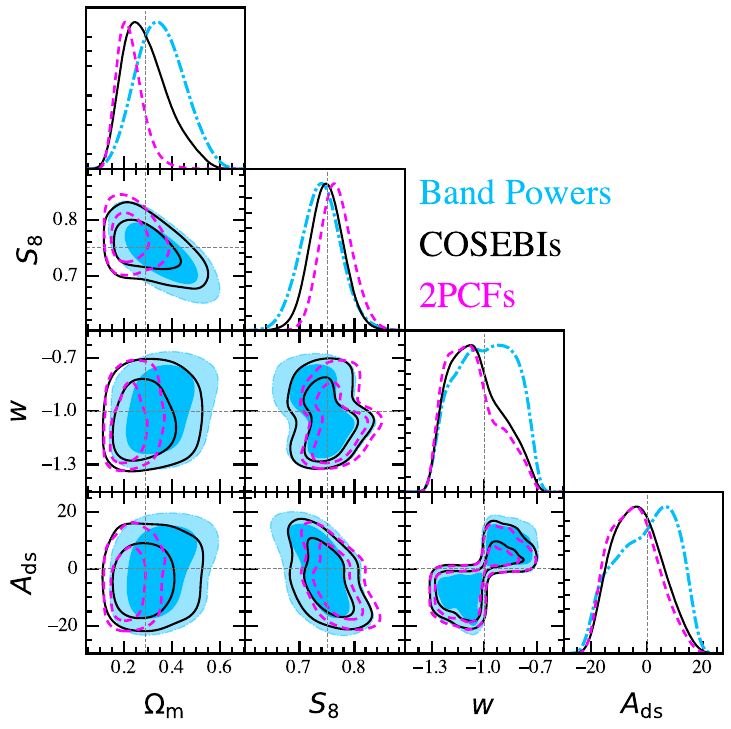}
  \caption{Constraints ($68 \%$ and $95 \%$ marginalised contours) on the key parameters $\Omega_{\rm m}$, $S_8$, $w$, and $A_{\rm ds}$ from all KiDS-1000 statistics sets: Contours for Band Powers (blue), COSEBIs (black) and Correlation Functions (magenta). The dashed lines represent COSEBIs mean values from the $\Lambda$CDM case. The DS parameter $A_{\rm ds}$ has units of $\rm [b/GeV]$.}
\label{Fig:K1K_DS_free_CMB}
\end{figure}

We thus analyze the BPs, COSEBIs and 2PCFs statistics and produce the posterior distribution for several parameters, including $\Omega_{\rm m}$, $S_8$, the DS parameter $A_{\rm ds}$, and $w$, as shown in \autoref{Fig:K1K_DS_free_CMB}. 
Note that although a key parameter, $w$, is prior-dominated, the KiDS-1000 data alone constrains the other parameter of the model to $\vert A_{\rm ds} \vert \lesssim 20$ $\rm [b/GeV]$ (at 68 $\%$ C.L.). This implies that the lensing data alone exhibits sensitivity to the non-linear effects of the interaction. Furthermore, we report the mean (and the marginalised 68$\%$ confidence values) of the several parameters and log-Bayes factor values for DS and $w$CDM ($A_{\rm ds} = 0$) in reference to $\Lambda$CDM, ($w = -1, A_{\rm ds} = 0$), and it is computed by $\log_{10} K = \log_{10} \left(\mathcal{Z}_{\rm{model}}/ \mathcal{Z}_{\Lambda\mathrm{CDM}} \right)$, where $\mathcal{Z}_{\rm{model}}$ represents the evidence of the model. \autoref{tab:bestfit_K1K_params} summarizes those results for the complete KiDS-1000 sets of statistics. As seen by the obtained log-Bayes factors, none of the cosmological models exhibits a definitive advantage over the rest. 

\subsection{KiDS-1000 + CMB+BAO combined analysis}\label{subsec:with_cmb}

\begin{figure} 
    \centering
    \includegraphics[width=0.5\textwidth]{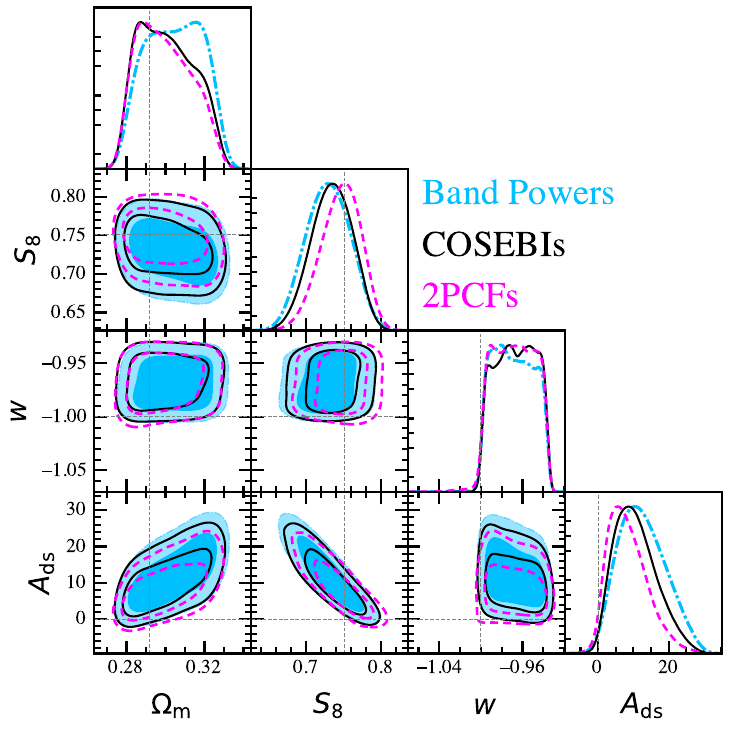}
  \caption{Constraints ($68 \%$ and $95 \%$ marginalised contours) on the key parameters $\Omega_{\rm m}$, $S_8$, $w$ and $A_{\rm ds}$ from all KiDS-1000 statistics sets combined with CMB+BAO: Contours for Band Powers (blue), COSEBIs (black) and Correlation Functions (magenta). The dashed lines represent COSEBIs mean values from the $\Lambda$CDM case. The units of $A_{\rm ds}$ are $\rm [b/GeV]$.}
\label{Fig:K1K_DS_CMB}
\end{figure}

As shown above, the KiDS-1000 data provides only an upper bound on the DS amplitude of $\vert A_{\rm ds} \vert \lesssim 20$ $\rm b/GeV$.
Aiming to constrain the model further, we supplement our analysis with additional information from the Planck measurements of the CMB temperature and polarisation~\citep{Planck:2018nkj}, as well as from BAO measurements from 6dFGS~\citep{Beutler:2011hx}, SDSS-MGS~\citep{Ross:2014qpa} and BOSS~\citep{BOSS:2016wmc}. 

In practice, we apply a prior on cosmological parameters derived from the posterior of the Planck TT,TE,EE+lowE+BAO analysis of the $w$CDM model\footnote{From \url{https://pla.esac.esa.int/pla/}}~\citep{Planck:2018vyg}, see \autoref{tab:priors_k1k_cmb_bao} for details. Despite this previous CMB+BAO analysis not including the effects of the dark sector interaction, it is a good approximation to use it for our combination. This is because we expect the CMB to be insensitive to the effects of $A_{\rm ds}$ ~\citep{Pourtsidou:2016ico}, since those only occur at late time; while the BAO is only sensitive to the expansion history, which is unaltered by Dark Scattering from $w$CDM. Still, to be conservative, we use instead flat priors on cosmological parameters taken from the 1D $2\,\sigma$ constraints of the CMB+BAO analysis.\footnote{We compared this setup against using Gaussian priors of the same width, but this did not result in differences in our posteriors.} This approximate method allows us to obtain a robust constraint of the $A_{\rm ds}$ parameter from the combination of KiDS-1000 with CMB and BAO data.

\begin{table}
  \centering 
  \begin{tabular}{|c|c|}
    \textbf{Parameter} & \textbf{Prior}  \\ \hline \hline $\omega_{\mathrm{b}}$   &  $\unif [0.022,  0.0226]$ \\ \hline 
    $\omega_{\mathrm{cdm}}$ &  $\unif [0.1174, 0.1223]$ \\ \hline
    $h$  & $\unif [0.6594, 0.7163]$ \\ \hline
    $n_s$ & $\unif [0.9571, 0.9736]$ \\ \hline
    $\ln(10^{10} A_s)$ & $\unif [3.0131, 3.0765]$\\ \hline
    $w$  & $\unif [-1.1591, -0.9347]$  \\ \hline
    $m_\nu$  & Fixed   \\ \hline
    $|A_{\rm ds}|$   &  $\unif [0.0, 30.0]$  \rm b/GeV \\ \hline
    $c_{\rm min}$ & $\unif [2, 4]$  \\ \hline
    $\eta_0$ & Derived \\ 
    \hline
    \hline
    \end{tabular}
   \caption{
We report the setup of the priors considered to cosmological parameters, which are sourced from the Planck TT,TE,EE+lowE+BAO analysis of the $w$CDM model with extended bounds to $2\,\sigma$.}
\label{tab:priors_k1k_cmb_bao}
\end{table}

The results of this analysis are shown in \autoref{Fig:K1K_DS_CMB} and \autoref{tab:bestfit_K1K_params_CMB_priors}. As reference, in Appendix \ref{app:A} we show the full contour plot of this analysis.
The posteriors show a clear preference for values for $A_{\rm ds} > 0$ for all KiDS-1000 statistics, and consequently $w > -1$. In particular, we see in the $w-A_{\rm ds}$ contour, an approximately $2\,\sigma$ deviation from $\Lambda$CDM, with the COSEBIs analysis giving $A_{\rm ds} = 10.6^{+4.5}_{-7.3}\ \rm b/GeV$. This contrasts with the KiDS-only result for which there was no preference for a non-zero value of the interaction strength. Note also that while the $w$ constraint appears prior-dominated, the CMB+BAO information is only enforcing it to be in the range $w\in [-1.159, -0.935]$, which only accounts for the upper bound on $w$, and then it is representing a data-driven constraint. The lower bound on $w$ is instead given by the physical prior enforcing the equal signs of $A_{\rm ds}$ and $(1+w)$. It is therefore the substantial preference for a positive interaction amplitude that is driving $w$ to the region of $w>-1$.

The physical explanation for this result is as follows. As we have already seen in \autoref{Fig:interaction_term_b}, a positive $A_{\rm ds}$ value represents a suppression in amplitude of the matter power spectrum due to an additional frictional force (see \autoref{Eq:interaction_eom}). This in turn decelerates the collapse of dark matter density fluctuations, reducing structure formation at late times. Since we essentially fixed the primordial amplitude by using CMB information in this analysis, the preference for a low late-time amplitude (i.e. the $S_8$ tension) is converted into a preference for a positive $A_{\rm ds}$. This effect is also evident in the anti-correlation of $A_{\rm ds}$ with the $S_8$ parameter displayed in \autoref{Fig:K1K_DS_CMB}.

\begin{figure}   
\includegraphics[width=0.5\textwidth]{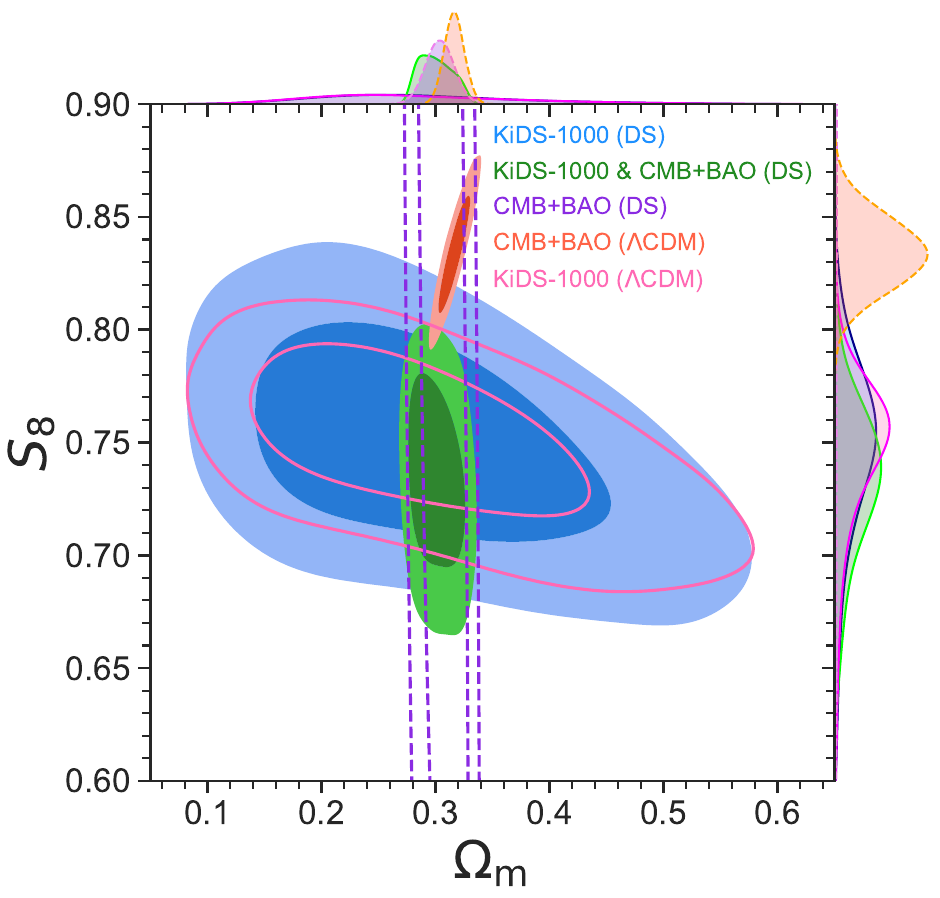}
\caption{Constraints projected on the $\Omega_{\rm m}$-$S_8$ plane ($68 \%$ and $95 \%$ C.L.). Firstly, we show in dashed violet an estimation of the constraints from the CMB+BAO analysis alone for the DS model, while in solid filled orange the full CMB+BAO analysis for the $\Lambda$CDM model. The filled blue contour represents DS constraints from the KiDS-1000 (COSEBIs only) analysis, while in solid pink lines we show the $\Lambda$CDM constraints. Finally, in green the constraints from KiDS-1000 \& CMB+BAO joint analysis for the DS model is displayed.}
\label{Fig:cosebis_all_models}
\end{figure}

\autoref{Fig:cosebis_all_models} shows the comparison of contours on the $\Omega_{\rm m}$-$S_8$ plane for the DS analysis with KiDS-1000 data as well as the KiDS-1000+CMB+BAO joint analysis (both evaluating the COSEBIs statistics). An approximation of the analysis of CMB+BAO (excluding KiDS-1000) for DS is also included. To obtain this contour we re-scaled the $S_8$ values of the $w$CDM TT,TE,EE+lowE+BAO chains by the DS growth factor for a broad range of values of $A_{\rm ds}$ (again assuming no constraining power from CMB+BAO on $A_{\rm ds}$, as justified above). This results in a complete broadening of the constraints in the $S_8$ direction, illustrating that, in the DS model, the CMB does not constrain the late-time amplitude, since it is insensitive to one of the parameters controlling it -- the interaction amplitude $A_{\rm ds}$. Cosmic shear then constrains the late-time amplitude and together with CMB+BAO, determines the value of $A_{\rm ds}$ that resolves the tension between early and late Universe probes of the amplitude of density fluctuations.\footnote{Note that the additional small difference between the CMB+BAO and the KiDS+CMB+BAO contours is due to the differences in the priors employed. More specifically, the flat priors employed in the KiDS-1000 scenarios, do not exactly correspond to the correlated near-Gaussian posteriors inherent in the CMB+BAO estimation.} 

For comparison, we include also in \autoref{Fig:cosebis_all_models} the results for the $\Lambda$CDM model from CMB+BAO as well as those from KiDS-1000. We can see that between $\Lambda$CDM and DS, the $S_8$ constraint is broadened, given the additional amplitude parameter being fitted. Additionally, when the CMB+BAO information is added, we can see that the $S_8$ constraint shifts to lower values. This is because $A_{\rm ds}$ is constrained to be positive, which can only lower the amplitude. This is clear when comparing with the $\Lambda$CDM case, as they have similar upper bounds on $S_8$ (corresponding to $A_{\rm ds}=0$), but differ in the lower bounds.

\begin{table}
\centering
\begin{tabular}{|c|c|c|c|}
  & \multicolumn{1}{c}{\textbf{Band Powers}}                       & \multicolumn{1}{c}{\textbf{COSEBIs}}                           & \multicolumn{1}{c}{\textbf{2PCFs}} \\
  & \multicolumn{1}{c}{\textbf{(CMB+BAO)}}                       & \multicolumn{1}{c}{\textbf{(CMB+BAO)}}                           & \multicolumn{1}{c}{\textbf{(CMB+BAO)}} \\
  \hline 
    $S_{8}$  
    & $0.729 \pm 0.029$ 
    & $0.734 \pm 0.027$
    & $0.746^{0.029}_{-0.024}$
    \\    \hline
    $A_{\rm ds}$  
    & $12.5^{+5.7}_{-7.8}$ $\rm b/GeV$ 
    & $10.6^{+4.5}_{-7.3}$ $\rm b/GeV$
    & $8.4^{+3.8}_{-6.7}$ $\rm b/GeV$
    \\     \hline
    $w$  
    & $-0.969^{+0.015}_{-0.026}$
    & $-0.967^{+0.027}_{-0.015}$
    & $-0.968 \pm 0.019$
    \\ \hline
    $c_{\rm min}$  
    & $2.97 \pm 0.50$ 
    & $2.43^{+0.15}_{-0.39}$
    & $2.36^{+0.12}_{-0.34}$
    \end{tabular}
    \caption{Mean and marginalised 68$\%$ constraints on key parameters of the DS model and the baryonic parameter, from the combined analysis of all KiDS-1000 probes with CMB+BAO measurements. The results are also presented in \autoref{Fig:K1K_DS_CMB}. Note that the constraints on $w$ are also prior dominated for each probe.
    }
\label{tab:bestfit_K1K_params_CMB_priors}
\end{table}

\section{Summary and Conclusion}
\label{sec:conclusion}

In this paper we constrained the Dark Scattering model using the $\sim 1000 \ \text{deg}^2$ galaxy shear catalogue from the KiDS survey. In order to enhance the efficiency of the inference pipeline in our analysis, we implemented emulators for the matter power spectrum, both at linear and non-linear order, with the latter built upon the framework of the halo model reaction. Furthermore, we have trained an emulator for baryonic feedback, a correction also included in the power spectrum. The emulators developed for this analysis will be publicly accessible at \code{DS-emulators}.

Our results show that the KiDS-1000 data constrains the DS parameter to be $\vert A_{\rm ds} \vert \lesssim 20$ $\rm b/GeV$ at 68\% C.L., as displayed in \autoref{Fig:K1K_DS_free_CMB}. Thus, we interpret that the KiDS-1000 cosmic shear catalog is sensitive to a combination of the growth history of this IDE model (i.e. the redshift evolution of $\sigma_8(z)$) and its specific non-linear effects.

In the joint analysis of KiDS-1000 with CMB and BAO information we obtain a stronger constraint on the DS model parameter, finding now $A_{\rm ds} = 10.6^{+4.5}_{-7.3}$ $\rm b/GeV$ (for COSEBIs, see \autoref{Fig:K1K_DS_CMB} for all cases). We find that the combined analysis favors positive values of $A_{\rm ds}$, since these lead to a reduction of the amplitude of the matter power spectrum at late times, as illustrated in \autoref{Fig:interaction_term_b}. While in the KiDS-only case, this reduction could be compensated by an increase of the primordial amplitude, the CMB and BAO information essentially breaks that degeneracy, thus allowing for a clearer determination of the interaction amplitude $A_{\rm ds}$. Accordingly, \autoref{Fig:cosebis_all_models} shows that the DS model is consistent with early and late-Universe measurements, thus offering a viable approach to alleviate the $S_8$ tension linked to the measured value of $A_{\rm ds}$.

\noindent It is important to emphasize the benefit of emulator-based methodologies employed in this paper. Implementing emulators from \code{CosmoPower}, we efficiently obtained the contours for the KiDS-1000 pipeline within a mere few minutes of computation, in a 24 CPU cores machine. 

With this computational tool at hand, we have at reach the analysis of different weak lensing probes, such as DES-Y3, or the recent joint analysis of KiDS-1000 + DES-Y3, with the purpose of constraining the DS model. These emulators can also be readily used for analyses including photometric galaxy clustering in a 3x2pt analysis, and can also be easily extended to constrain the DS model with a time-dependent $w(z)$. Additionally, these emulators will allow for an accelerated exploitation of the stage-IV data that will soon become available from Euclid and Rubin's LSST. There, the importance of this fast analysis tool is even more critical, given the huge nuisance-parameter spaces that need to explored. Moreover, the implications of adopting alternative models for intrinsic alignments (IA), baryonic feedback and other systematics can also be more efficiently explored in the future with these emulators. This would then allow for a thorough investigation of their impact on the inference of the DS parameter and its degeneracies with e.g. baryonic feedback, or massive neutrinos.

\section*{Acknowledgments}

We are grateful to Alejandro Aviles for helpful suggestions during the student conference at Instituto de Ciencias Físicas (ICF) and Sebastien Fromenteau for insightful feedback on this work. We are grateful to Anna Porredon for useful discussions.
We acknowledge use of the Cuillin computing cluster of the Royal Observatory, University of Edinburgh, and the Chalcatzingo machine provided by ICF, UNAM. KC acknowledges support from a CONAHCyT studentship. ASM acknowledges support from the MSSL STFC Consolidated Grant ST/W001136/1. AP is a UK Research and Innovation Future Leaders Fellow [grant MR/X005399/1]. PC's research is supported by grant RF/ERE/221061. JCH acknowledges support from program UNAM-PAPIIT, grant IG102123 “Laboratorio de Modelos y Datos (LAMOD) para proyectos de Investigaci\'on Cient\'{\i}fica: Censos Astrof\'{\i}sicos”.
Based on observations made with ESO Telescopes at the La Silla Paranal Observatory under programme IDs 177.A-3016, 177.A-3017, 177.A-3018 and 179.A-2004, and on data products produced by the KiDS consortium. The KiDS production team acknowledges support from: Deutsche Forschungsgemeinschaft, ERC, NOVA and NWO-M grants; Target; the University of Padova, and the University Federico II (Naples).
For the purpose of open access, the author has applied a Creative Commons Attribution (CC BY) licence to any Author Accepted Manuscript version arising from this submission. This work made use of publicly available software. We acknowledge the usage of \code{GetDist} \citep{Lewis19} package for plotting contours of posteriors. The plotting style was provided from \cite{SciencePlots}.


\section*{Data Availability}
\phantomsection
\label{data_av}

The data underlying this article will be shared on reasonable request to the corresponding author. The public repositories of codes employed in this paper are the following:

\begin{itemize}
     \item \code{DS-CLASS} \github{https://github.com/PedroCarrilho/class_public/tree/IDE_DS} was used to obtain the DS linear spectrum. 
    \item \code{ReACT-IDE} \github{https://github.com/PedroCarrilho/ReACT/tree/react_with_interact} includes the DS modification.
    
    \item \code{HMCode} \github{https://github.com/alexander-mead/HMcode} provided us the DM-only pseudo power spectrum.

    \item \code{CosmoPower} \github{https://github.com/alessiospuriomancini/cosmopower.git} was the tool for training our emulators.
 
    \item \code{DS-emulators} \github{https://github.com/karimpsi22/DS-emulators.git} has available our emulators.

    \item  \code{Montepython} \github{https://github.com/brinckmann/montepython_public} performed our statistical analysis. 

    \item  \code{Multinest} \github{https://github.com/JohannesBuchner/MultiNest} was the sampler.

\end{itemize} 


\bibliographystyle{mnras}
\bibliography{mybib} 

\appendix
\section{Complementary results}
\label{app:A}

\begin{figure*} 
    \centering
    \includegraphics[width=\textwidth]{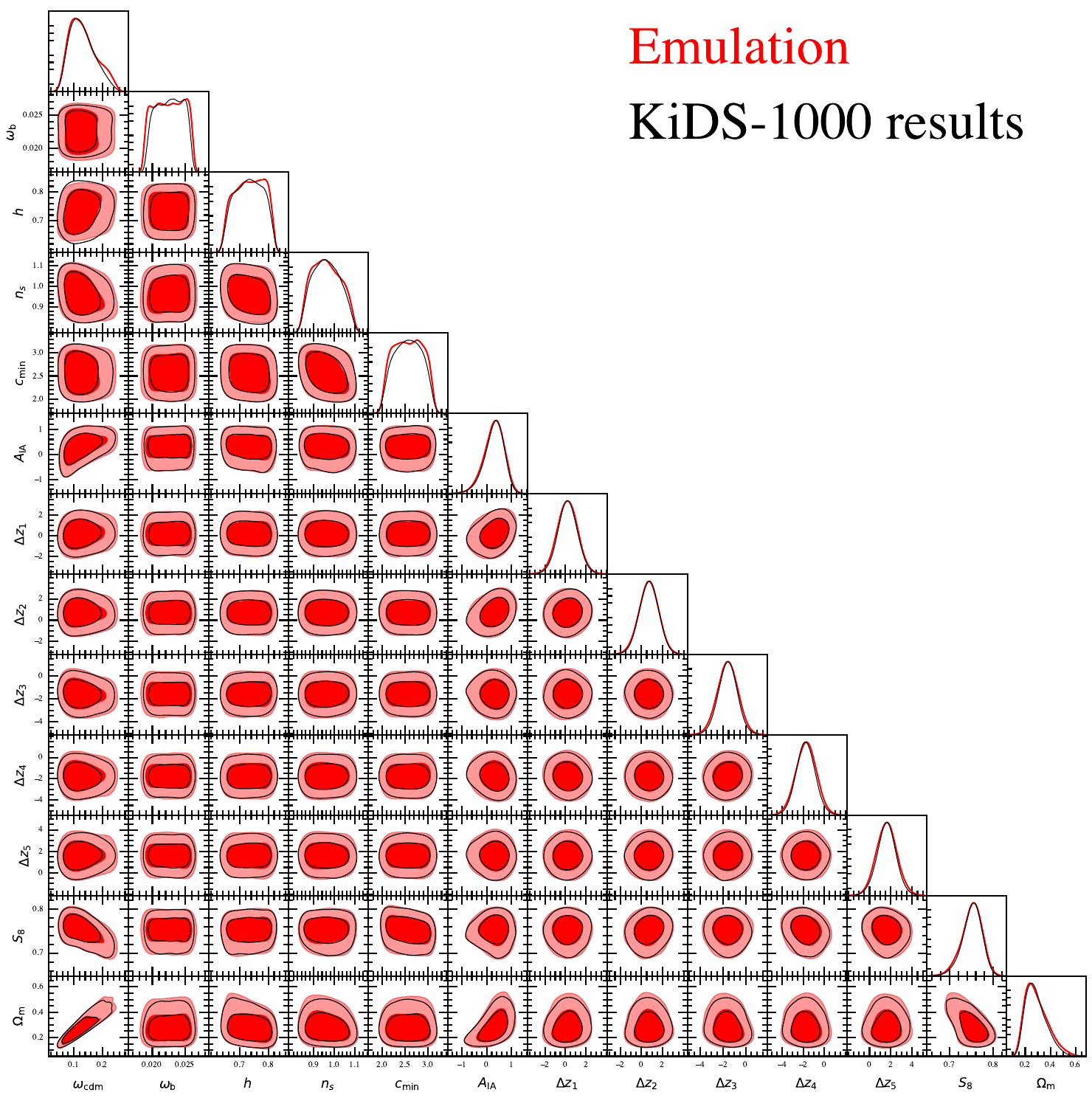}
  \caption{Full COSEBIs statistics posteriors from the KiDS-1000 chains \citep{KiDS:2020suj} (black lines), and from emulators with parameters fixed to reflect the $\Lambda$CDM case (solid red). The plot clearly shows the desired consistency of the emulated data with the official set. Note that the full emulator chains were calculate in a few minutes, while the sole Boltzmann solver (CAMB) takes days to process this data.}
\label{Fig:LCDM_emu_vs_k1k_data}
\end{figure*}

In this appendix we present important supplementary contour plots of this work. The first validity test for emulators involved assessing whether they are capable to reproduce the KiDS-1000 $\Lambda$CDM official analysis \citep{KiDS:2020suj}. This is shown in \autoref{Fig:LCDM_emu_vs_k1k_data}, which displays a full comparison between the publicly available KiDS-1000 results for $\Lambda$CDM and posteriors obtained through our emulators in a few a minutes.  

\begin{figure*} 
    \centering
\includegraphics[width=\textwidth]{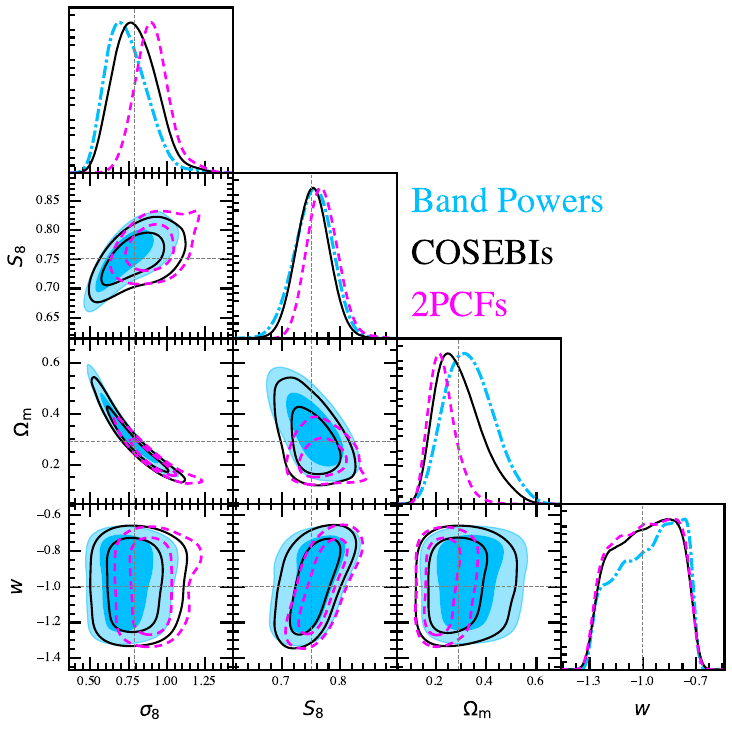}
  \caption{Posteriors from the emulator, for the $w$CDM cosmology, using Band Powers (blue), COSEBIs (black) and Two point Correlation Functions (magenta) statistics. The dashed lines represent COSEBIs mean values from $\Lambda$CDM case.
  }
\label{Fig:wCDM_emu_all_sets}
\end{figure*}

A second test has been to evaluate the emulators in the context of the $w$CDM cosmology as reported in \citet{KiDS:2020ghu}. This is presented in  \autoref{Fig:wCDM_emu_all_sets} in which we maintain the same \code{MultiNest} settings.

\begin{figure*} 
    \centering
    \includegraphics[width=\textwidth]{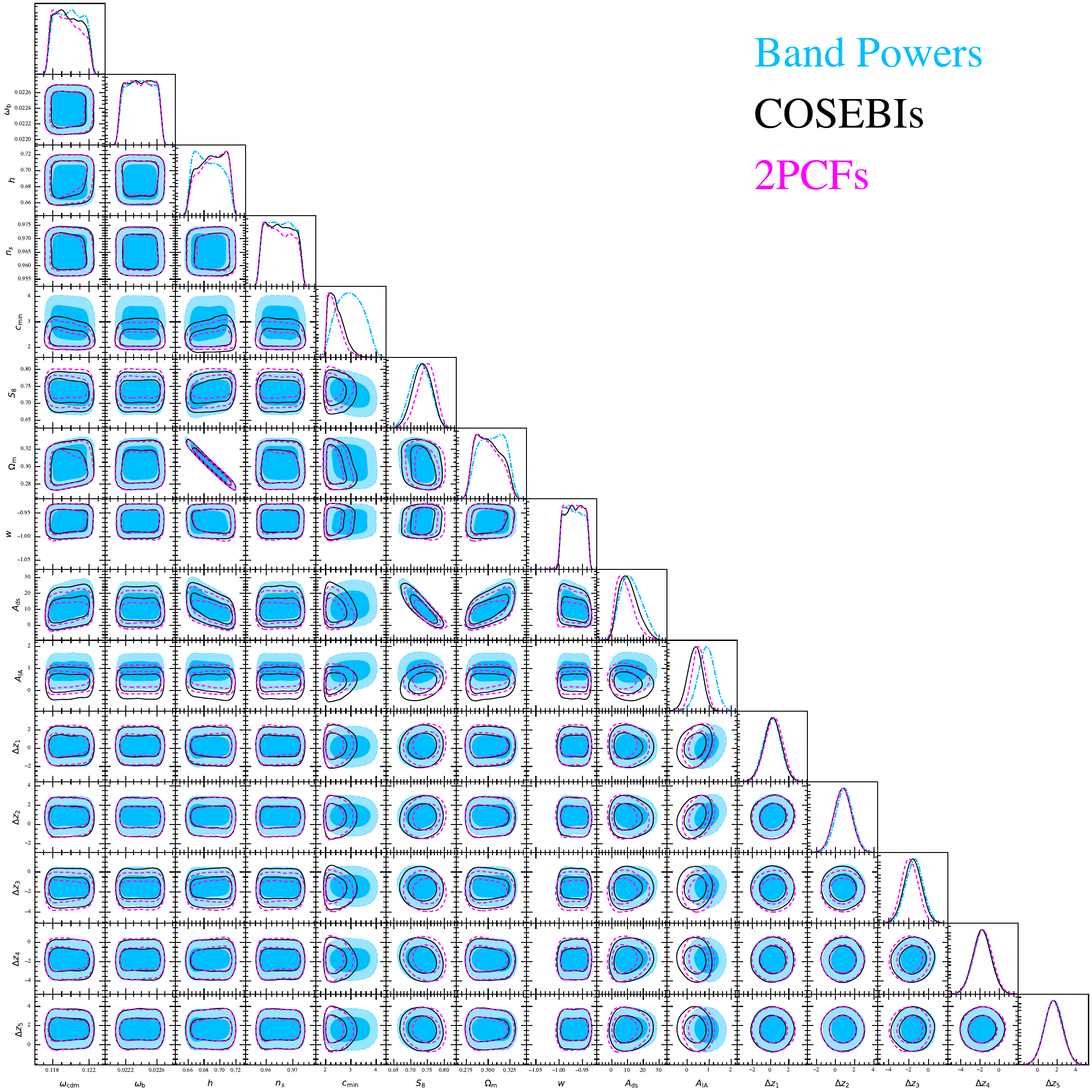}
  \caption{Full posteriors of DS model and also photo-z errors from KiDS-1000 and CMB+BAO combined analysis. The units of $A_{\rm ds}$ are $\rm [b/GeV]$.}
\label{Fig:all_k1k_sets}
\end{figure*}

Lastly, \autoref{Fig:all_k1k_sets} illustrates the posterior distributions of nine cosmological parameters and five nuisance parameters related to the redshift distributions ($\Delta z_i$). These distributions are derived from our combined analyses with CMB+BAO and considering the three distinct KiDS-1000 statistical approaches: Band Powers, COSEBIs, and 2PCFs. Such chains were obtained by using \code{MultiNest}.

\bsp	
\label{lastpage}
\end{document}